\documentclass[prb,preprint]{revtex4-1} 

\usepackage{amsmath}  
\usepackage{amsfonts} 
\usepackage{graphicx} 

\begin{document}


\title{An introduction to the New SI}

\author{Sandra Knotts}
\email{sknotts@pvsd.org}
\affiliation{Perkiomen Valley High School, Collegeville, Pennsylvania 19426}
\author{Peter J. Mohr}
\email{mohr@nist.gov}
\affiliation{National Institute of Standards and Technology,
Gaithersburg, Maryland 20899}
\author{William D. Phillips}
\email{wphillips@nist.gov}
\affiliation{National Institute of Standards and Technology
Gaithersburg, Maryland 20899}
\affiliation{University of Maryland, College Park, Maryland 20742}


\begin{abstract}

Plans are underway to redefine the International System of Units (SI)
around 2018.  The New SI specifies the values of certain physical
constants to define units.  This article explains the New SI in a way
that could be used to present it to high-school physics classes.

\end{abstract}

\maketitle 

\section{Introduction}

Physical science is based on measurements, and the results of
measurements are expressed in terms of units. For example, someone can
measure the length of a table and report that it is 1.4 m long, where m
stands for meters. This statement provides useful information to other
people, because there is general agreement about what a meter is. In
fact, there is a treaty among 55 nations, including the US, that
codifies agreement on exactly what the meter is. Not surprisingly, the
treaty is called the {\it Convention of the Meter}, and it also
specifies other units.\cite{bipm} According to the treaty, the
International System of Units, known as the SI, is agreed to be the
present day standard by which all participating member states set their
units.  Even though a majority of people in the US still use units such
as inches and pounds, the official standards for these units are linked
to the SI. For example, the US definition of the inch is that it is
exactly 0.0254 m.

Although the practice of establishing standards for measurements dates
back thousands of years, the {\it Convention of the Meter} was only
established in 1875 with seventeen nations initially signing on,
including the US.  Incidentally, the anniversary of the signing, 20 May,
is now known as World Metrology Day.   The SI, established in 1960, is
still more recent and continues to change`.\cite{bipmsi}  In fact, it is
anticipated that the definitions of measurement standards specified by
the SI will undergo a substantial change in
2018.\cite{2011191,2014050,newell}  In the New SI, units are defined by
assigning specific values to a set of physical
constants.\cite{2010054,2012158} Despite the fact that the definitions
will be changed, the effect on everyday measurements will be
imperceptible.  This paper describes the new SI and shows how units will
be based on assigned values of the constants.

\section{A short history of length}

The significance of the new SI, in which the units are defined by
defining some fundamental constants of nature, can be understood in part
by recalling the history of the unit of length.

In antiquity the unit of length was often tied to the dimensions of the
human body.  A ``foot'' was the length of a foot, a ``hand'' was the width
of a hand, a ``cubit'' was the length from the elbow to the fingertip plus
a hand, and so forth.  The convenience of such an arrangement is
obvious--the standard of length is always available and easy to use.  The
difficulty is also obvious--the standard was not really standard, varying
with each individual.  Various approaches attempted to remedy this
difficulty:  The foot was the length of the king's foot, or the average
of the lengths of some number of people's feet.  Such stratagems helped,
but did not eliminate the problem.  

The ancient Egyptians took an approach that is decidedly modern.
Building the pyramids demanded precision and consistency in measurement.
Some ancient metrologist established a ``primary standard'' cubit in the
form of a rod of a certain length (possibly tied to the dimensions of
the body of some chosen person).  Artisans who did the actual cutting
and sizing of stones used a ``working standard'' of length that was
compared regularly to the primary standard.  In this way the Egyptian
engineers guaranteed a consistent measure of length, protecting the
primary standard from the wear and tear of everyday use, while providing
easy access to the standard by transferring its accuracy to the working
standards in the field.  

Merchants of medieval Europe used a similar approach.  Towns would embed
a standard of length into a masonry wall in the town square, and all who
did business in that town could confirm that goods sold by length were
accurately represented.  Variations from town to town could, however,
cause problems.  Other approaches seem even more modern:  The inch was
defined as the length of three barleycorns.  This connection to a
universal feature of nature had an obvious appeal, but variations in
weather from one growing season to the next, as well as variations among
different plants, made such a standard less than perfect.  

The French revolution brought an opportunity for both scientific and
political reforms to the question of length.  A standard that was both
unchanging and universally available was taken to be the size of the
earth itself.  The meter was defined as a part in ten million of the
distance from the equator to the pole, through Paris.  The difficulty of
surveying even the accessible part of the pole-to-equator arc was great
enough that the legal standard of length became the ``meter of the
archives,'' a platinum rod whose end-to-end length was the closest to
the meter based on the surveyed distance.  

Eventually the meter and the metric system as a whole became more widely
and internationally accepted.  An international agreement in 1875, the
Treaty of the Meter, established the International Bureau of Weights and
Measures and the intention of creating a new international standard of
length, the International Prototype Meter (IPM).  This was a
platinum-iridium bar with fine lines (scratches) between whose centers
was one meter.  This ``line'' standard had an obvious advantage over the
earlier ``end'' standard of the archives in that using it did not wear
down its length.  Copies of this bar, with calibrations of their
differences from the IPM, were distributed to the signatories of the
Treaty.  

Now, the entire world had a system for measuring length analogous to
that of the ancient Egyptians.  But problems soon arose.  The distance
between the scratches on meter bars could be compared with an accuracy
on the order of one or two tenths of a micrometer, that is, about a part
in $10^7$.  The ability to compare wavelengths of light and to compare
distances measured in terms of such wavelengths outstripped the
measurement of lengths based on the distance between two scratches on a
metal bar.  From early in the 20$^{\rm th}$ century metrologists juggled two
systems of length measurement: the legal system based on the meter bar
and a (more precise) {\it de facto} one based on the wavelength of light from
a cadmium lamp.  

This uncomfortable situation was rectified in 1960 when the
international community agreed on a new definition of the legal (SI)
meter based on the wavelength of light from a krypton lamp.  Now, the
best metrological techniques of optics were used for determining
lengths.  But the seeds of the destruction of even this advanced
definition were already in the wind.  1960 saw the first demonstration
of a laser, and it soon was clear that laser light could be more ``pure''
having a narrower spread of wavelengths than could a krypton lamp.
Furthermore, the distribution of krypton wavelengths was found to be
asymmetric, making the definition of the meter ambiguous.  The result
was that, as in the early 20th century, metrologists of the late 20th
century began to use a {\it de facto} standard of length based on a He-Ne
laser locked to the wavelength of a particular transition in the I$_2$
molecule.  Many thought that a new definition would follow, based on the
new {\it de facto} standard, as it had in 1960.  But a better idea emerged,
one that was a harbinger of the current reform of the SI.  

In 1983, the international community adopted a definition of the meter
as the distance that light travels in a certain length of time.
Effectively, this defines the meter by defining $c$, the speed of light.
One of the beauties of this definition, in contrast to the earlier
definitions of the meter in terms of the cadmium or krypton wavelengths,
is that it makes no choice of an atom or molecule as the standard.
Anyone with a good laser, and a means of measuring the frequency $f$ of
the light it emits, automatically has a length standard graduated in
wavelengths $\lambda$ of light according to the expression $\lambda =
c/f$.  This beautiful definition makes the standard of length available
to everyone at all times.  If better lasers and better ways of measuring
their frequencies become available, the same definition will embody
those improvements.  There will be no need to repeat the sad history of
the 20$^{\rm th}$ century's length definition, switching from Pt-Ir bars
to Cd to Kr to I$_2$.  

Today we are reforming the SI in a manner analogous to the beautiful
definition of the meter.  By defining Planck's constant, the electron
charge, the Boltzmann constant and Avogadro's number, we bring the power
of the 1983 definition of the speed of light to the rest of the SI.  At
the same time, we eliminate {\it de facto} definitions of the Volt and
the Ohm that until now forced us to use two systems of units, the SI and
the {\it de facto} ``practical'' system, for electrical quantities.  In
the 19$^{\rm th}$ century James Clerk Maxwell dreamed of a system of
measurement in which the invariant constants of nature would provide the
standards.  Today we realize that dream. 

\section{The defining constants}

The concept of defining units by assigning values to physical constants
requires some explanation. The previous example of a table that is 1.4 m
long illustrates this idea. The fact that the table is 1.4 m long is
determined by comparing its length to the length of an object that is 1
m long, in particular a meter stick.  Now if through some natural
disaster all of our meter sticks disappeared, we would have a problem
determining lengths. However if the table survived the disaster, we
would have a physical object by which we could now define the meter. We
could say the meter is the length standard so that if we measure the
table it will come out to be 1.4 m long. In this way we're taking a
physical object, assigning a numerical value to its length, and this in
turn determines the meter, the unit of length. 

The definition of the new SI is based on this principle. By assigning a
value to the result of a measurement of a property of nature, one can
define the unit that is used to make the measurement.  Unlike the table,
which is a particular physical object, the properties of nature, or
quantities that are actually used for the definitions, are expected to
be unchanging and available to anybody with the appropriate measurement
tools.

\subsection{Frequency and time}

The frequency standard for the SI is established by assigning a value to
the frequency of a particular atomic transition.  It is currently, and
for a while in the New SI will continue to be, the frequency
corresponding to the transition between the two hyperfine levels in the
ground state of a cesium atom.  Here frequency refers to the frequency
of the electromagnetic radiation absorbed or emitted when the atom makes
a transition between the two hyperfine levels.  The value assigned to
the frequency is $9\,192\,631\,770$ cycles per second, where the unit
cycles per second is called hertz or Hz.  This value, about 9 GHz (the G
stands for $10^9$), is in the microwave frequency range, similar to
kitchen microwave oven frequencies, which can be about 2.5 GHz.

Once this standard of frequency is defined, it is possible to determine
other frequencies by comparison to this standard or determine a second
by counting cycles of the frequency.  There are commercial instruments
that can count periodic cycles at this frequency.  Since there are
exactly $9\,192\,631\,770$ cycles of the radiation in a second, we can
time one second by counting off that number of cycles of the standard
frequency.

It is this latter possibility that gives the name to atomic clocks.  The
actual functioning of the cesium clock is as a frequency standard, but
it can be used to measure time by counting cycles of electromagnetic
radiation.

\subsection{Velocity and length}

Microwaves and light are forms of electromagnetic radiation that differ
only in their frequencies. The frequency of light is roughly 500 THz
(the T stands for $10^{12}$).  Despite the difference in frequency
between microwaves and light, all electromagnetic radiation travels at
the same speed, even if it is measured by someone moving relative to the
source of the radiation. This is the principle of Einstein's theory of
special relativity.  This speed is just the velocity of light, denoted
by $c$, which has an assigned value of $c = 299\,792\,458$ meters per
second in both the current SI and the New SI. This provides a universal
velocity standard to which other velocities may be compared.

At the same time this provides a standard for length in the SI. The
wavelength of light, denoted by $\lambda$, is the distance between peaks
of the electromagnetic radiation. For light traveling past a point in
space, the number of wavelengths that go by in a second is the
frequency, denoted by $f$, of the radiation. So the total distance
traveled by light in one second is the wavelength times the number of
wavelengths that pass a stationary point, or the wavelength times the
frequency.  This distance is just $299\,792\,458$ m because the velocity
of light is that many meters per second.  This is expressed by the
equation $\lambda f = c$.  Because of this relationship, for light of a
given frequency the wavelength will be a certain known length which can
be used to determine the length of one meter. In practice, one way of
doing this is by using an interferometer which allows distances to be
determined by counting off wavelengths as the distance between two
mirrors is changed.

\subsection{Voltage and resistance}

Modern measurements of electrical quantities are made using two quantum
phenomena from condensed matter physics, namely the Josephson effect and
the quantum Hall effect.

For measurements of voltage, the Josephson effect is used and yields a
value in terms of an applied frequency (which is defined as described
earlier) and the Josephson constant, denoted by $K_{\rm J}$, which is
about $484$ THz/V.  Resistance measurements are made with the quantum
Hall effect and are expressed in terms of the von Klitzing constant,
denoted by $R_{\rm K}$, given by about $25\,813~{\rm\Omega}$, where the
unit is ohms.  Since voltage and resistance can be measured in terms of
these two constants, it follows from Ohm's law that current can also be
determined by measuring the voltage across a resistor of known
resistance when the current flows through.

Presently, both the Josephson and von Klitzing constants are accurately,
but not exactly, known.  On the other hand, the measurements of voltage
and resistance in terms of these constants can be made even more
accurately than the constants themselves are presently known in SI
units.  In the New SI, these constants will for all practical purposes
be exactly known.  The values are assigned as described in the next
section.

\subsection{Charge and action}

The unit of electrical charge, denoted by $e$, is the charge of a proton
or the negative of the charge of an electron.  It is approximately $e =
1.6\times10^{-19}$ C, where the unit is coulombs or C.  In the New SI,
the unit of charge is assigned an exact value.

The Planck constant, denoted by $h$, is a universal constant associated
with quantum mechanics and is approximately $6.6\times10^{-34}$ J s.  It
has units of action, which is equivalent to energy multiplied by time.
This constant has a number of roles in quantum physics.  It is the
constant of proportionality that gives the energy of a photon, a quantum
of electromagnetic radiation, in terms of its frequency $f$; the energy
of a photon is $E = hf$.  The Planck constant is also the unit of
angular momentum.  The smallest angular momentum of an elementary
particle such as the electron is $h/4\pi$.  It happens that in the
present SI, action and angular momentum have the same units.  The Planck
constant also appears in the Heisenberg uncertainty relation.  In the
New SI, the Planck constant is assigned an exact value.

An additional role of the Planck constant is in the Josephson effect and
the quantum Hall effect.  Both the Josephson constant and the von
Klitzing constants are simple functions of the unit electric charge and
the Planck constant.  The relations are $K_{\rm J} = 2e/h$ and $R_{\rm
K} = h/e^2$.

Since in the New SI, both the electric charge and the Planck constant
are assigned exact values, this defines the units of charge C and action
J s.  As an additional consequence, because of the relation to the
Josephson and von Klitzing constants, both of those constants are also
exactly known, and electrical measurements as described in the previous
section will give results in terms of exactly known constants.

The assigned values of $e$ and $h$ that will be used for the
redefinition are not yet completely set.  The exact values will depend
on whatever relevant experimental information is available at the time
of the redefinition.

\subsection{Mass and power}
\label{sec:map}

In the SI, an important unit is the kilogram for mass.  In the New SI,
the mass of an object in kilograms can be determined from the units
already defined.  Once the unit of charge and the Planck constant are
defined, electrical units in terms of the Josephson and von Klitzing
constants are known, and this is enough to determine the mass of an
object using these units and others previously defined.  

The principle can be understood through a simple thought experiment to
measure mass.  Suppose we construct an apparatus consisting of an
electric motor that can wind up a string that has a mass hanging at the
end.  If the motor is run to lift the mass, it will work against the
force of gravity on the mass.  The motor will exert a certain average
power which is the increase in potential energy of the mass divided by
the time it takes to lift it.  The necessary power is proportional to
the mass being lifted.  The same average electrical power is consumed by
the motor lifting the mass, which is the voltage times the current being
supplied to the motor.  Since these quantities can be measured with the
electrical units already defined, the mass can be determined.

The local acceleration of gravity is also needed in order to know the
force of gravity on the mass.  It can be determined by dropping an
object and measuring its acceleration with the time and velocity units
already defined.

In this way, comparing mechanical power to electrical power, mass can be
measured.  The most precise way of doing this is using an apparatus
called the watt balance, where the watt is the unit of power.  The watt
balance is essentially an electrical scale that measures mass by
weighing it in the presence of gravity.  To avoid friction, which would
be a problem for the simple motor example mentioned, the watt balance
makes the measurement in two steps.  One step is the measurement of the
current through a coil in a magnetic field that supports a mass with no
motion, and the other step is the determination of the magnetic field by
moving the coil through it and measuring the induced voltage on the
coil.  The result is a combination of electrical quantities that is
equivalent to power, which is the source of the name of the apparatus.
Model watt balances are described by \citet{quinn13} and
\citet{schlamminger}.

Electrical power, the product $VI = VV^\prime/R$, where the current is
determined by measuring the voltage $V^\prime$ across a resistor with
resistance $R$ calibrated with the quantum Hall effect, is proportional
to the combination $K_{\rm J}^2R_{\rm K} = 4/h$.  The electric charge
cancels from this combination, so it does not need to be exactly known
in order to measure mass.  The mass determination depends only on $h$
and the time and length standards mentioned earlier.  However, having
$e$ also exact allows for separate measurement of both voltage and
resistance.

Another method of determining mass in the New SI is to produce an object
with a mass of one kilogram as a standard.  This can accurately be done
based on the fact that in the New SI, because the Planck constant is
exact, the mass of one silicon-28 atom is known in kilograms with a
precision of about 10 significant figures from experiments that measure
the recoil of an atom when it absorbs a photon, for example.  A
spherical single crystal of these atoms can be made with the suitable
number of atoms needed to make up one kilogram.  The number of atoms in
the spherical crystal is determined by the measured volume of the sphere
and the measured average spacing between the atoms in the crystal
structure.  This provides a method of producing an object with a mass of
one kilogram that is independent of the watt balance.

\subsection{Amount of substance} \label{sec:mas}

The amount of stuff, or more precisely, amount of substance in the SI is
just the number of items, atoms, molecules, {\it etc.}, under
consideration.  The unit of amount of substance is the mole, which
consists of approximately $6.02\times 10^{23}$ items.  In the New SI,
this number is fixed by assigning an exact value to the Avogadro
constant, denoted by $N_{\rm A}$, which is defined to be that number of
items per mole.  As in the case of the elementary charge and the Planck
constant, the exact number will be determined by the results of
experiments done up to the time of the redefinition so that it is close
to the number in the current SI.

\subsection{Temperature}

Temperature can be measured with a gas thermometer based on the ideal
gas law: $pv = nRT$.  In this equation, $p$ is the pressure of the gas
in a container, $v$ is the volume of the container, $n$ is the number of
moles of gas in the container, $R$ is the ideal gas constant, and $T$ is
the temperature of the gas.  One can imagine that the gas chamber is in
contact with the object whose temperature is being measured, so the
temperature of the gas will eventually be equal to the temperature to be
measured.  The temperature $T$ of the gas may be found by determining
the rest of the terms in the ideal gas law equation and solving for $T$.

This can be done, based on the definitions already mentioned, as
follows.  The pressure $p$ is force per unit area.  A standard for force
can be calibrated by comparison to the force of gravity on a known mass,
and area is just length squared.  The volume $v$ is just length cubed.
The number of moles $n$ of gas in the chamber is the total number of
atoms $N$ divided by the Avogadro constant $n = N/N_{\rm A}$.  This
leaves the problem of determining the ideal gas constant $R$, which is
related to the Boltzmann constant $k$ by $R = N_{\rm A}k$.  In the New
SI, the Boltzmann constant is one of the exactly defined constants,
which means that $R$ is also exactly known.  So, using this definition,
we get the temperature from the various other measured aspects of the
gas thermometer.

Note that from the fact that $nR = Nk$, it can be seen that the Avogadro
constant is not necessary for such a measurement of temperature,
although it provides a more convenient measure of the amount of gas than
the number of atoms does.

The Boltzmann constant is approximately $1.38\times10^{-23}$ J/K, where
K or kelvin is the unit of temperature in the SI.  Evidently, the
product $kT$ has units of energy.  In fact, other methods of measuring
temperature also involve measuring energy and the combination $kT$ is
what is measured.  In this sense, the Boltzmann constant is a conversion
factor from energy to temperature, which gives a number for temperature
which is easier to communicate than the energy would be.

\subsection{Light}

For light, the candela, abbreviated cd, is the unit of luminous
intensity, which is the power emitted by a light source in a particular
direction multiplied by a factor $K_{\rm cd}$, which takes into account
the sensitivity of the eye to various colors of light.

One candela is approximately the luminous intensity observed for a
candle, where it gets its name.  If the source has a luminous intensity
that is the same in all directions, then the total luminous flux is
$4\pi$ lumen.

In the New SI, the candela is defined by specifying that the factor
$K_{\rm cd}$ for green light at 540 THz is exactly 683 lumen per watt.

\section{New SI}

The previous sections should make it clear how the specification of the
values of seven constants defines the New SI units.  This is the basis
for the New SI, which together with the definitions of the various units
with special names, such as the Joule, in terms of the units defined by
the constants, is sufficient to define the entire SI.  The International
Committee for Weights and Measures intends to propose a revision of the
SI as follows:\cite{res1}

The International System of Units, the SI, will be the system of units
in which:
\begin{itemize}
\item
the ground state hyperfine splitting frequency of the cesium 133 atom
$\Delta \nu(^{133}{\rm Cs})_{\rm hfs}$ is exactly $9\,192\,631\,770$
hertz,
\item
the speed of light in vacuum $c$ is exactly $299\,792\,458$ meter per
second,
\item
the Planck constant $h$ is exactly $6.626\ldots\times10^{-34}$ joule
second,
\item
the elementary charge $e$ is exactly $1.602\ldots\times10^{-19}$ coulomb,
\item
the Boltzmann constant $k$ is exactly $1.380\ldots\times10^{-23}$ joule
per mole,
\item
the Avogadro constant $N_{\rm A}$ is exactly $6.022\ldots\times10^{23}$
reciprocal mole,
\item
the luminous efficacy $K_{\rm cd}$ of monochromatic radiation of
frequency $540\times 10^{12}$ Hz is exactly 683 lumen per watt,
\end{itemize}

\noindent where the dots $(\ldots)$ will be replaced by additional
digits some of which have yet to be determined by experiments to be done
up to the time of the redefinition.  From then on, the constants will be
defined to be exactly those values.

\section{New SI vs. current SI}

The essential change from the current SI to the New SI is in the
definitions of the units.  In the current SI, seven base units are
defined by the following statements:

\begin{itemize}
\item
The meter is the length of the path travelled by light in vacuum during
a time interval of $1/299\,792\,458$ of a second.
\item
The kilogram is the unit of mass; it is equal to the mass of the
international prototype of the kilogram.
\item
The second is the duration of $9\,192\,631\,770$ periods of the
radiation corresponding to the transition between the two hyperfine
levels of the ground state of the cesium 133 atom.
\item
The ampere is that constant current which, if maintained in two straight
parallel conductors of infinite length, of negligible circular
cross-section, and placed 1 meter apart in vacuum, would produce between
these conductors a force equal to $2\times10^{-7}$ newton per meter of
length.
\item
The kelvin, unit of thermodynamic temperature, is the fraction
$1/273.16$ of the thermodynamic temperature of the triple point of
water.
\item
\begin{enumerate}
\item
The mole is the amount of substance of a system which contains as
many elementary entities as there are atoms in 0.012 kilogram of carbon
12; its symbol is ``mol''.
\item
When the mole is used, the elementary entities must be specified and
may be atoms, molecules, ions, electrons, other particles, or specified
groups of such particles.
\end{enumerate}
\item
The candela is the luminous intensity, in a given direction, of a source
that emits monochromatic radiation of frequency $540\times10^{12}$ hertz
and that has a radiant intensity in that direction of $1/683$ watt per
steradian.
\end{itemize}

In the New SI, no units are singled out as base units.  The concept of
base units is a carryover from when the meter and kilogram were both
defined in terms of metal objects.  In the New SI, it is recognized that
the definitions are not directly linked to the earlier base units in a
one-to-one relationship.  For example, in the New SI, the speed of light
$c$ is given a particular value and that defines the combination of
units m/s.  Similarly, giving $h$ a particular value defines the
combination J s or kg m$^2/$s.  The defining constants provide
definitions of enough combinations of units so that they are all
uniquely determined.\cite{2008027}

Although the forms of all of the definitions are different, some of the
units are unchanged in the New SI.  In particular, the meter, second,
and candela are equivalent in the New and current SI.  On the other
hand, the kilogram, ampere, kelvin, and mole are defined differently.
In fact, in the current SI, the kilogram is still the mass of a metal
artifact kept in a vault near Paris, as it was first in 1889.

\section{Why change?}

The New SI is a significant improvement over the current SI, which is
the reason for changing.  Some of the improvements are as follows.

The current definition of the kilogram, in terms of a metal artifact, is
problematic, because the mass of the artifact is changing relative to
the mass of similar copies.  In addition, it needs to be washed
before being used for measurements, and in that process, the mass also
changes.  These variations are larger than the uncertainty in the
measurement of mass using a watt balance or a silicon crystal sphere for
a comparison standard, so the New SI will allow for more reliable mass
measurements.

Electrical measurements using the Josephson effect and the quantum Hall
effect are more accurate than the Josephson and von Klitzing constant
are currently known in terms of SI units.  As a result, since 1990,
electrical measurements have been made in terms of assigned values for
these constants.  The assigned values are not directly tied to the SI,
so electrical measurements are not actually done in terms of SI units.
With the exact specification of the electric charge and Planck constant
in the New SI, the Josephson and von Klitzing constants are well
defined, and the results of electrical measurements will be given in
actual SI units.

The new definition of the mole states the number of entities in a mole.
In the current SI, the mole is the number of carbon atoms in 0.012 kg of
carbon, but the actual number is not given, which is less clear.

For temperature, the kelvin is currently defined in terms of the triple
point of water, that is the unique temperature at which water coexists
as ice, liquid and vapor.  However, this property of water depends on
the purity and isotopic composition of the water used. The kelvin is
better defined if linked to an exact numerical value of the Boltzmann
constant $k$.

When expressed in terms of the units of the New SI, the values of many
physical constants have smaller uncertainties, and some, besides the
defining constants, are exact.

\section{Conclusion}

It is expected that the SI will be redefined as described here
approximately in 2018.  The time for this to happen depends mainly on
experiments that determine the value of the Planck constant being
sufficiently accurate that the defined value in the New SI will be
correct.

The New SI is another step in the improvements in measurement standards
that have been happening over the millennia.

\section{Topics for further study}

Some topics related to the redefinition that could be the basis for
further studies are:\cite{2010054,2012158}
\begin{itemize}
\item
History of units.
\item
Determination of physical constants.
\item
Hyperfine splitting frequency.
\item
Interferometry.
\item
Josephson effect.
\item
Quantum Hall effect.
\item
Watt balance experiment.
\item
Crystal lattice spacing measurements.
\item
Atom recoil experiments.
\item
Temperature measurements.

\end{itemize}

\appendix

%
%
%
%
%
%

%

\end{document}